\documentclass[12pt,a4paper]{article}
\usepackage{amsmath}
\usepackage[dvips]{graphicx}
\usepackage{times}
\begin{document}
\begin{titlepage}
\title{On the ridge-like  structures  in the nuclear and hadronic  reactions}
\author{S.M. Troshin, N.E. Tyurin\\[1ex]
\small  \it Institute for High Energy Physics,\\
\small  \it Protvino, Moscow Region, 142281, Russia}
\normalsize
\date{}
\maketitle

\begin{abstract}
 We  briefly comment on the ridge-like structure origin in the nuclear and hadronic  reactions emphasizing that
this structure in the two-particle correlation function can result from 
the rotation of the transient state of matter. 
\end{abstract}
\end{titlepage}
\setcounter{page}{2}

It is well known that the studies of multiparticle production in hadron and nucleus processes  provide  a clue to the mechanisms of
deconfinement and hadronization. The deconfined state found at RHIC
reveals the properties of the perfect liquid, being strongly
interacting collective state \cite{rhic}.

Nowadays, when experimental program of the LHC has started , it is important to analyze newly obtained experimental 
data and try to make  first conclusions on the nature of a  matter produced in
$pp$  collisions, i.e. is it  weakly interacting or it remains to be a strongly interacting one
as it was observed at RHIC in $AA$ collisions? In the latter case one can expect that proposed in \cite{intje}
mechanism related to the rotation of transient matter should be working at the LHC energies and 
therefore  the observed at RHIC phenomena should be observed in $pp$-collisons also.

The ridge structure was observed first at RHIC in peripheral  collisons of nuclei in the two-particle 
correlation function in the near-side
jet production (cf. recent paper \cite{rhicr} and references therein). It was demonstrated that the ridge particles have
 a narrow $\Delta\phi$ correlation distribution (where $\phi$ 
is an asimuthal angle) and wide $\Delta\eta$ correlations ($\eta$ is a pseudorapidity). The ridge phenomenon was associated
with the collective effects of a medium. 

The similar structure in the two-particle correlation function was observed by
the CMS Collaboration \cite{ridgecms}. This is rather surprising result because the ridge structure was observed for the first time
in $pp$--collisions. Those collisions are commonly treated as  the kind of ``elementary'' ones under the heavy-ion studies 
and therefore often used  as the reference
process for detecting deconfined phase formation on the base of difference between $pp$- and $AA$-collisions. 
It is evident now that such approach should be revised in view of this new and  
unexpected  experimental result.

The particle
production machanism  proposed in the model  \cite{intje} takes into  account  the 
geometry  of the overlap region and dynamical properties of
the transient state in hadron interaction. This picture assumes  deconfinement at the initial stage of interaction.
The transient state appears as a rotating medium of massive quarks and pions which hadronize and 
form multiparticle final state.  Essential point for this 
rotation is the non-zero impact parameter in the collision.

Indeed, the inelastic overlap function $h_{inel}(s,b)$,

\[
h_{inel}(s,b)\equiv\frac{1}{4\pi}\frac{d\sigma_{inel}}{db^2},
\]
has a peripheral impact parameter dependence at the energy $\sqrt{s}=7$ TeV due to the 
reflective scattering \cite{intja}.
Note, that unitarity equation rewritten at high energies
for the elastic amplitude $f(s,b)$ has the form
\[
\mbox{Im} f(s,b)=h_{el}(s,b)+ h_{inel}(s,b)
\]
and $h_{inel}(s,b)$ is the sum of all inelastic channel contributions.
Due to this peripherality, the mean multiplicity 
\[
 \langle n\rangle (s)=\frac{\int_0^\infty bdb  \langle n\rangle (s,b) h_{inel}(s,b)}
{\int_0^\infty bdb h_{inel}(s,b)}
\]
gets the main contribution from the collisions with non-zero impact parameters.
Thus, one can assume that the events with high multiplicity at the LHC energy $\sqrt{s}=7$ TeV  correspond
to the peripheral hadron collisions \cite{intja}. Thus, at the LHC energy $\sqrt{s}=7$ TeV there is a dynamical
selection of peripheral region in impact parameter space responsible for the inelastic processes. In the nuclear reactions
 such selection is provided by the relevant  experimental adjustments. Note, that  the ridge-like structure in the
nuclear reactions has also been observed in peripheral collisions only \cite{rhicr}.

The  geometrical picture of hadron collision at non-zero impact parameters
implies that the generated massive
virtual  quarks in overlap region  could obtain very large initial orbital angular momentum
at high energies. Due to strong interaction
between quarks this orbital angular momentum  leads to the coherent rotation
of the quark system located in the overlap region as a whole  in the
$xz$-plane (Fig. 1). This rotation is similar to the liquid rotation
where strong correlations between particles momenta exist. 
Thus, the orbital angular momentum should be realized  as a coherent rotation
of the quark-pion liquid  as a whole. 
The assumed particle production mechanism at moderate transverse
momenta is an excitation of  a part of the rotating transient state of  massive constituent
quarks (interacting by pion exchanges). 
\begin{figure}[h]
\begin{center}
\resizebox{7cm}{!}{\includegraphics*{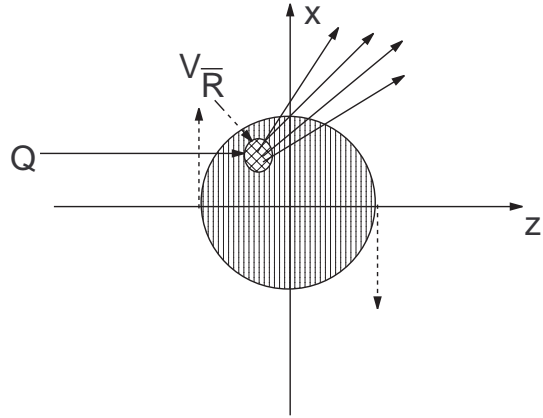}}
\caption{\small \it {Interaction of the constituent quark with rotating quark-pion liquid.}}
\end{center}
\end{figure}
Due to the fact that the transient matter is strongly interacting, the excited parts
should be located closely  to the periphery  of the rotating transient state otherwise absorption
 would not allow to quarks and pions  leave the interaction region (quenching). 

The mechanism is sensitive
 to the particular  direction of rotation and to the rotation plane orientatation. This will
lead to the narrow distribution of the two-particle correlations in $\Delta\phi$. However,
two-particle correlation could have broad distribution in polar angle ($\Delta\eta$) in the above mechanism (Fig. 1).
 Quarks in the exited part of the cloud
could have different values of the two components of the momentum (with its third component
lying in the rotation  $xz$-plane) since the exited region $V_{\bar R}$ has significant extension. 

Thus,  the ridge-like structure observed  in the high multiplicity events by the CMS Collaboration 
can be an experimental manifestation of the coherent rotation of the transient matter in hadron
collisions. The narrowness of the two-particle correlation distribution in the asimuthal angle is the
distinctive feature of this mechanism. 

There should be other experimentally
observed effects of this collective effect, one of them is the directed flow $v_1$ in hadron reactions,
with fixed impact parameter discussed in \cite{intje}.  Rotation of transient matter
will affect also elliptic flow  $v_2$ and  average transverse momentum of secondary particles produced in proton-proton collisions \cite{mpla}.
Due to  rotation the density of  massive quarks will be
 different in the different parts of the rotating cloud, it will be smaller in the central part and bigger at
 the peripheral part of cloud due to the centrifugal effect. At the same time the quarks
 in the peripheral part have a maximal transverse momenta and therefore we should observe correlation
 of the multiplicity and transverse momentum.
It would lead, in particular, 
to the following relation of the average transverse momentum with the mean multiplicity of secondaries
\[
 \langle p_T\rangle(s)=a+b\langle n\rangle (s).                                                                                              
\]
This relation is in a good agreement with experimental data \cite{mpla}.

The above discussion shows that the nature of the state of matter 
revealed at the LHC in proton collisions is the same as the nature of the state revealed
  at RHIC in nuclei collisions.

\end{document}